\documentstyle[iopconf1]{article}

\def\WIMPZILLA{{\sc wimpzilla}}
\def\WIMPZILLAS{{\sc wimpzillas}}
\def\WIMP{{\sc wimp}}
\def\WIMPS{{\sc wimps}}
\def\LTE{{\sc lte}}

\input epsf

\begin{document}

\title{WIMPZILLAS!}

\author{
Edward W.\ Kolb\dag\ddag\footnote{E-mail: rocky@rigoletto.fnal.gov}, \\
Daniel J.\ H.\ Chung\S\footnote{E-mail: 
           djchung@feynman.physics.lsa.umich.edu}, \\
Antonio Riotto\P\footnote{E-mail: riotto@nxth04.cern.ch}
}

\affil{\dag\ Theoretical Astrophysics \\Fermi National Accelerator
Laboratory\\ Batavia, Illinois 60510}

\affil{\ddag\ Department of Astronomy and Astrophysics \\ 
Enrico Fermi Institute \\
The University of Chicago\\ Chicago, Illinois 60637}

\affil{\S\ Department of Physics \\ The University of Michigan \\
Ann Arbor, Michigan 48109}

\affil{\P\ Theory Division\\ CERN\\ CH-1211 Geneva 23, Switzerland}

\beginabstract 
There are many reasons to believe the present mass density of the
universe is dominated by a weakly interacting massive particle
(\WIMP), a fossil relic of the early universe.  Theoretical ideas and
experimental efforts have focused mostly on production and detection
of {\it thermal} relics, with mass typically in the range a few GeV to
a hundred GeV.  Here, I will review scenarios for production of {\it
nonthermal} dark matter.  Since the masses of the nonthermal \WIMPS\
are in the range $10^{12}$ to $10^{16}$ GeV, much larger than the mass
of thermal wimpy \WIMPS, they may be referred to as \WIMPZILLAS.  In
searches for dark matter it may be well to remember that ``size does
matter.''
\endabstract

\section{Introduction}

\addtocounter{footnote}{-3}

\def\simlt{\stackrel{<}{{}_\sim}}
\def\simgt{\stackrel{>}{{}_\sim}}
\def\question#1{{{\marginpar{\small \sc #1}}}}
\def\ktilde{{{\tilde{k}}}}
\def\etatilde{{{\tilde{\eta}}}}

There is conclusive evidence that the dominant component of the matter
density in the universe is dark.  The most striking indication of the
existence of dark matter is the dynamical motions of astronomical
objects.  Observations of flat rotation curves for spiral galaxies
\cite{halodm} indicates that the dark component of galactic halos is
about ten times the luminous component.  Dynamical evidence for DM in
galaxy clusters from the velocity dispersion of individual galaxies,
as well as from the large x-ray temperatures of clusters, is also
compelling \cite{clusterdm}.  Bulk flows, as well as the peculiar
motion of our own local group, also implies a universe dominated by
dark matter \cite{velocitydm}.

The mass of galaxy clusters inferred by their gravitational lensing of
background images is consistent with the large dark-to-visible mass
ratios determined by dynamical methods \cite{lensingdm}.

There is also compelling evidence that the bulk of the dark component
must be nonbaryonic.  The present baryonic density is restricted by
big-bang nucleosynthesis to be less than that inferred by the methods
discussed above \cite{bbndm}.  The theory of structure formation from
the gravitational instability of small initial seed inhomogeneities
requires a significant nonbaryonic component to the mass density
\cite{lssdm}.

In terms of the critical density, $\rho_C=3 H_0^2 M_{{\rm Pl}}^2/8\pi
=1.88\times10^{-29}$g cm$^{-3}$ with Hubble constant $H_0\equiv 100 h$
km\,sec$^{-1}$Mpc$^{-1}$ and Planck mass $M_{{\rm Pl}}$, the
dark-matter density inferred from dynamics is $\Omega_{{\rm DM}}\equiv
\rho_{{\rm DM}}/\rho_C\simgt 0.3$. In addition, the most natural
inflation models predict a flat universe, {\it i.e.,} $\Omega_0=1$,
while standard big-bang nucleosynthesis implies that ordinary baryonic
matter can contribute at most $10\%$ to $\Omega_0$.  This means that
about $90\%$ of the matter in our universe may be dark.

\section{Thermal Relics---Wimpy WIMPS}

It is usually assumed that the dark matter consists of a species of a
new, yet undiscovered, massive particle, traditionally denoted by $X$.
It is also often assumed that the dark matter is a thermal relic, {\it
i.e.,} it was in chemical equilibrium in the early universe.

A thermal relic is assumed to be in local thermodynamic equilibrium
(\LTE) at early times.  The {\it equilibrium} abundance of a particle,
say relative to the entropy density, depends upon the ratio of the
mass of the particle to the temperature.  Define the variable $Y\equiv
n_X/s$, where $n_X$ is the number density of WIMP $X$ with mass $M_X$,
and $s \sim T^3$ is the entropy density.  The equilibrium value of
$Y$, $Y_{EQ}$, is proportional to $\exp(-x)$ for $x\gg 1$, while
$Y_{EQ}\sim$ constant for $x\ll 1$, where $x=M_X/T$.

\begin{figure}[t]
\centering
\leavevmode\epsfxsize=300pt  \epsfbox{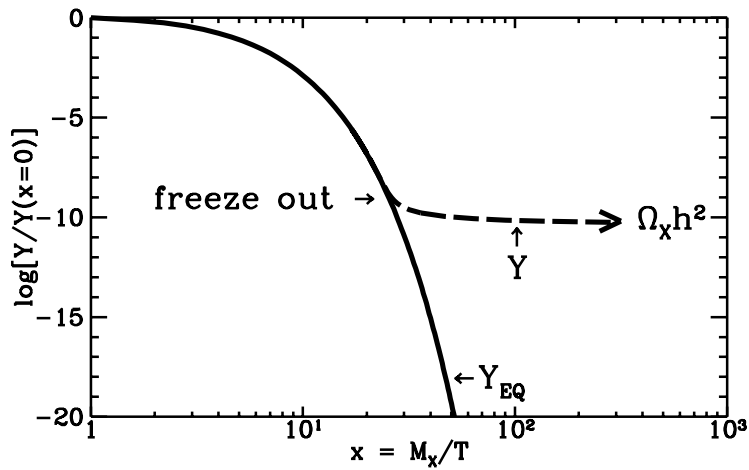}
\caption{\label{thermal} A thermal relic starts in \LTE\ at $T\gg
M_X$.  When the rates keeping the relic in chemical equilibrium become
smaller than the expansion rate, the density of the relic relative to
the entropy density freezes out.}
\end{figure}

A particle will track its equilibrium abundance as long as reactions
which keep the particle in chemical equilibrium can proceed rapidly
enough.  Here, rapidly enough means on a timescale more rapid than the
expansion rate of the universe, $H$.  When the reaction rate becomes
smaller than the expansion rate, then the particle can no longer track
its equilibrium value, and thereafter $Y$ is constant.  When this
occurs the particle is said to be ``frozen out.''  A schematic
illustration of this is given in Fig.\ \ref{thermal}.

The more strongly interacting the particle, the longer it stays in
\LTE, and the smaller its eventual freeze-out abundance.  Conversely,
the more weakly interacting the particle, the larger its present
abundance.  The freeze-out value of $Y$ is related to the mass of the
particle and its annihilation cross section (here characterized by
$\sigma_0$) by \cite{book}
\begin{equation}
Y \propto \frac{1}{M_X m_{Pl} \sigma_0} \ .
\end{equation}
Since the contribution to $\Omega$ is proportional to $M_Xn_X$, which
in turn is proportional to $M_XY$, the present contribution to
$\Omega$ from a thermal relic roughly is {\em independent} of its
mass,\footnote{To first approximation the relic dependence depends
upon the mass only indirectly through the dependence of the
annihilation cross section on the mass.}  and depends only upon the
annihilation cross section.  The cross section that results in
$\Omega_Xh^2\sim 1$ is of order $10^{-37}$cm$^2$, of the order the
weak scale.  This is one of the attractions of thermal relics.  The
scale of the annihilation cross section is related to a known mass
scale.

The simple assumption that dark matter is a thermal relic is
surprisingly restrictive.  The largest the annihilation cross section
can be is roughly $M_X^{-2}$.  This implies that large-mass \WIMPS\
would have such a small annihilation cross section that their present
abundance would be too large.  Thus one expects a maximum mass for a
thermal WIMP, which turns out to be a few hundred TeV
\cite{griestkam}.

The standard lore is that the hunt for dark matter should concentrate
on particles with mass of the order of the weak scale and with
interaction with ordinary matter on the scale of the weak force. This
has been the driving force behind the vast effort in dark matter
direct detection described in this meeting by Cabrera \cite {blas},
Liubarsky \cite{liubarsky}, Bernabei \cite{rita}, Ramachers
\cite{ramachers}, and Baudis \cite{laura}.

In view of the unitarity argument, in order to consider {\it thermal}
\WIMPZILLAS, one must invoke, for example, late-time entropy
production to dilute the abundance of these supermassive particles
\cite{k}, rendering the scenario unattractive.

\section{Nonthermal Relics---WIMPZILLAS}

There are two necessary conditions for the \WIMPZILLA\ scenario.
First, the \WIMPZILLA\ must be stable, or at least have a lifetime
much greater than the age of the universe.  This may result from, for
instance, supersymmetric theories where the breaking of supersymmetry
is communicated to ordinary sparticles via the usual gauge forces
\cite{review}. In particular, the secluded and the messenger sectors
often have accidental symmetries analogous to baryon number. This
means that the lightest particle in those sectors might be stable and
very massive if supersymmetry is broken at a large scale \cite{raby}.
Other natural candidates arise in theories with discrete gauge
symmetries \cite{discrete} and in string theory and M theory
\cite{john,dimitridark98}.

It is useful here to note that \WIMPZILLA\ decay might be able to
account for ultra-high energy cosmic rays above the
Greisen--Zatzepin--Kuzmin cutoff \cite{vadimvaleri,subircr}.  A wimpy little
thermal relic would be too light to do the job, a \WIMPZILLA\ is needed.

The second condition for a \WIMPZILLA\ is that it must not have been in
equilibrium when it froze out ({\it i.e.,} it is not a thermal relic),
otherwise $\Omega_Xh^2$ would be much larger than one.  A sufficient
condition for nonequilibrium is that the annihilation rate (per
particle) must be smaller than the expansion rate: $n_X\sigma|v|<H$,
where $\sigma |v|$ is the annihilation rate times the M{\o}ller flux
factor, and $H$ is the expansion rate.  Conversely, if the dark matter
was created at some temperature $T_*$ {\it and} $\Omega_Xh^2<1$, then
it is easy to show that it could not have attained equilibrium.  To
see this, assume $X$'s were created in a radiation-dominated universe
at temperature $T_*$.  Then $\Omega_Xh^2$ is given by
\begin{equation}
\Omega_Xh^2 = \Omega_\gamma h^2(T_*/T_0)m_Xn_X(T_*)/\rho_\gamma(T_*)\ , 
\end{equation}
where $T_0$ is the present temperature.  Using the fact that
$\rho_\gamma(T_*) = H(T_*) M_{Pl} T_*^2$, $n_X(T_*)/H(T_*) =
(\Omega_X/\Omega_\gamma) T_0 M_{Pl} T_*/M_X$.  One may safely take the
limit $\sigma |v| < M_X^{-2}$, so $n_X(T_*)\sigma |v| / H(T_*)$ must
be less than $(\Omega_X / \Omega_\gamma) T_0 M_{Pl} T_* / M_X^3$.
Thus, the requirement for nonequilibrium is
\begin{equation}
\left( \frac{200\,{\rm TeV}}{M_X}\right)^2 \left( \frac{T_*}{M_X}
\right) < 1 \ .
\end{equation}
This implies that if a nonrelativistic particle with $M_X\simgt 200$
TeV was created at $T_*<M_X$ with a density low enough to result in
$\Omega_X \simlt 1$, then its abundance must have been so small that
it never attained equilibrium. Therefore, if there is some way to
create \WIMPZILLAS\ in the correct abundance to give $\Omega_X\sim 1$,
nonequilibrium is automatic.

Any \WIMPZILLA\ production scenario must meet these two criteria.
Before turning to several \WIMPZILLA\ production scenarios, it is
useful to estimate the fraction of the total energy density of the
universe in \WIMPZILLAS\ at the time of their production that will
eventually result in $\Omega\sim 1$ today.

The most likely time for \WIMPZILLA\ production is just after
inflation.  The first step in estimating the fraction of the energy
density in \WIMPZILLAS\ is to estimate the total energy density when
the universe is ``reheated'' after inflation.

Consider the calculation of the reheat temperature, denoted as
$T_{RH}$. The reheat temperature is calculated by assuming an
instantaneous conversion of the energy density in the inflaton field
into radiation when the decay width of the inflaton energy,
$\Gamma_\phi$, is equal to $H$, the expansion rate of the universe.

The reheat temperature is calculated quite easily \cite{book}.  After
inflation the inflaton field executes coherent oscillations about the
minimum of the potential.  Averaged over several oscillations, the
coherent oscillation energy density redshifts as matter: $\rho_\phi
\propto a^{-3}$, where $a$ is the Robertson--Walker scale factor.  If
$\rho_I$ and $a_I$ denotes the total inflaton energy density and the
scale factor at the initiation of coherent oscillations, then the
Hubble expansion rate as a function of $a$ is
\begin{equation}
H(a) = \sqrt{\frac{8\pi}{3}\frac{\rho_I}{M^2_{Pl}}
	\left( \frac{a_I}{a} \right)^3}\ .
\end{equation}
Equating $H(a)$ and $\Gamma_\phi$ leads to an expression for $a_I/a$.
Now if all available coherent energy density is instantaneously
converted into radiation at this value of $a_I/a$, one can define the
reheat temperature by setting the coherent energy density,
$\rho_\phi=\rho_I(a_I/a)^3$, equal to the radiation energy density,
$\rho_R=(\pi^2/30)g_*T_{RH}^4$, where $g_*$ is the effective number of
relativistic degrees of freedom at temperature $T_{RH}$.  The result
is
\begin{equation}
\label{eq:TRH}
T_{RH} = \left( \frac{90}{8\pi^3g_*} \right)^{1/4}
		\sqrt{ \Gamma_\phi M_{Pl} } \
       = 0.2 \left(\frac{200}{g_*}\right)^{1/4}
	      \sqrt{ \Gamma_\phi M_{Pl} } \ .
\label{eq:trh2}
\end{equation}
The limit from gravitino overproduction is $T_{RH} \simlt 10^{9}$ to
$10^{10}$ GeV.

Now consider the \WIMPZILLA\ density at reheating.  Suppose the
\WIMPZILLA\ never attained \LTE\ and was nonrelativistic at the time
of production.  The usual quantity $\Omega_X h^2$ associated with the
dark matter density today can be related to the dark matter density
when it was produced.  First write
\begin{equation} 
\frac{\rho_X(t_0)}{\rho_R(t_0)}
=\frac{\rho_X(t_{RH})}{\rho_R(t_{RH})}\:\left(\frac{T_{RH}}{T_0}\right),
\label{eq:transfromrh}
\end{equation}
where $\rho_R$ denotes the energy density in radiation, $\rho_X$
denotes the energy density in the dark matter, $T_{RH}$ is the reheat
temperature, $T_0$ is the temperature today, $t_0$ denotes the time
today, and $t_{RH}$ denotes the approximate time of
reheating.\footnote{More specifically, this is approximately the time
at which the universe becomes radiation dominated after inflation.} To
obtain $\rho_X(t_{RH})/\rho_R(t_{RH})$, one must determine when $X$
particles are produced with respect to the completion of reheating and
the effective equation of state between $X$ production and the
completion of reheating.

At the end of inflation the universe may have a brief period of matter
domination resulting either from the coherent oscillations phase of
the inflaton condensate or from the preheating phase
\cite{preheating}.  If the $X$ particles are produced at time
$t=t_{e}$ when the de Sitter phase ends and the coherent oscillation
period just begins, then both the $X$ particle energy density and the
inflaton energy density will redshift at approximately the same rate
until reheating is completed and radiation domination begins.  Hence,
the ratio of energy densities preserved in this way until the time of
radiation domination is
\begin{equation} 
\frac{\rho_X(t_{RH})}{ \rho_R(t_{RH})} \approx \frac{8\pi}{3}\:
\frac{\rho_X(t_{e})}{M_{Pl}^2 H^2(t_{e}) },
\end{equation} 
where $M_{Pl} \approx 10^{19}$ GeV is the Planck mass and most of the
energy density in the universe just before time $t_{RH}$ is presumed
to turn into radiation.  Thus, using Eq.\ \ref{eq:transfromrh}, one
may obtain an expression for the quantity $\Omega_X\equiv
\rho_X(t_0)/\rho_C(t_0)$, where $\rho_C(t_0)=3 H_0^2M_{Pl}^2/8\pi$ and
$H_0=100\: h$ km sec$^{-1}$ Mpc$^{-1}$:
\begin{equation} 
\Omega_X h^2 \approx \Omega_R h^2\:
\left(\frac{T_{RH}}{T_0}\right)\: 
\frac{8 \pi}{3} \left(\frac{M_X}{M_{Pl}}\right)\:
\frac{n_X(t_{e})}{M_{Pl} H^2(t_{e})}.
\label{eq:omegachi}
\end{equation}
Here $\Omega_R h^2 \approx 4.31 \times 10^{-5}$ is the fraction of
critical energy density in radiation today and $n_X$ is the density of
$X$ particles at the time when they were produced.

Note that because the reheating temperature must be much greater than
the temperature today ($T_{RH}/ T_0 \simgt 4.2 \times 10^{14}$), in
order to satisfy the cosmological bound $\Omega_X h^2 \simlt 1$, the
fraction of total \WIMPZILLA\ energy density at the time when they
were produced must be extremely small.  One sees from Eq.\
\ref{eq:omegachi} that $\Omega_Xh^2 \sim 10^{17}(T_{RH}/10^9\mbox
{GeV})(\rho_X(t_e)/\rho(t_e))$.  It is indeed a very small fraction of
the total energy density extracted in \WIMPZILLAS.

This means that if the \WIMPZILLA\ is extremely massive, the challenge
lies in creating very few of them.  Gravitational production discussed
in Section \ref{gravprod} naturally gives the needed suppression.
Note that if reheating occurs abruptly at the end of inflation, then
the matter domination phase may be negligibly short and the radiation
domination phase may follow immediately after the end of inflation.
However, this does not change Eq.\ \ref{eq:omegachi}.

\section{WIMPZILLA PRODUCTION}

\subsection{\label{gravprod} Gravitational Production}

First consider the possibility that \WIMPZILLAS\ are produced in the
transition between an inflationary and a matter-dominated (or
radiation-dominated) universe due to the ``nonadiabatic'' expansion of
the background spacetime acting on the vacuum quantum fluctuations
\cite{grav}.

The distinguishing feature of this mechanism is the capability of
generating particles with mass of the order of the inflaton mass
(usually much larger than the reheating temperature) even when the
particles only interact extremely weakly (or not at all) with other
particles and do not couple to the inflaton.  They may still be
produced in sufficient abundance to achieve critical density today due
to the classical gravitational effect on the vacuum state at the end
of inflation.  More specifically, if $0.04 \simlt M_X/H_I \simlt 2$,
where $H_I \sim m_\phi \sim 10^{13}$GeV is the Hubble constant at the
end of inflation ($m_\phi$ is the mass of the inflaton), \WIMPZILLAS\
produced gravitationally can have a density today of the order of the
critical density. This result is quite robust with respect to the
``fine'' details of the transition between the inflationary phase and
the matter-dominated phase, and independent of the coupling of the
\WIMPZILLA\ to any other particle.

Conceptually, gravitational \WIMPZILLA\ production is similar to the
inflationary generation of gravitational perturbations that seed the
formation of large scale structures.  In the usual scenarios, however,
the quantum generation of energy density fluctuations from inflation
is associated with the inflaton field that dominated the mass density
of the universe, and not a generic, sub-dominant scalar field.
Another difference is that the usual density fluctuations become
larger than the Hubble radius, while most of the \WIMPZILLA\
perturbations remain smaller than the Hubble radius.

There are various inequivalent ways of calculating the particle
production due to interaction of a classical gravitational field with
the vacuum (see for example \cite{fulling}, \cite{birrelldavies}, and
\cite{chitre}). Here, I use the method of finding the Bogoliubov
coefficient for the transformation between positive frequency modes
defined at two different times.  For $M_X/H_I \simlt 1$ the results
are quite insensitive to the differentiability or the fine details of
the time dependence of the scale factor.  For $0.04 \simlt M_X/H_I \simlt
2$, all the dark matter needed for closure of the universe can be made
gravitationally, quite independently of the details of the transition
between the inflationary phase and the matter dominated phase.

\begin{figure}[p]
\centering
\leavevmode\epsfxsize=300pt  \epsfbox{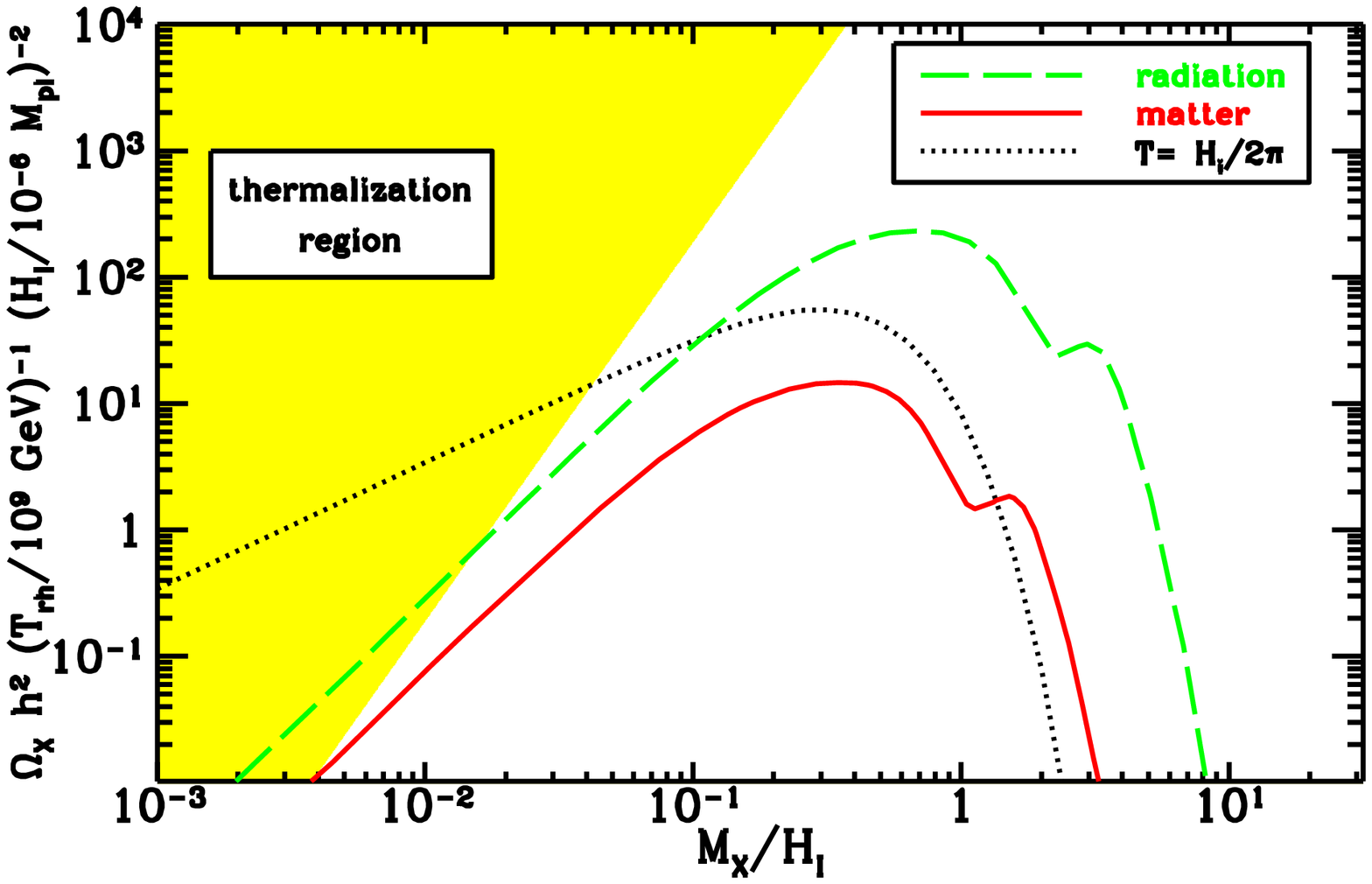}
\caption{\label{hawking} The contribution of gravitationally produced
\WIMPZILLAS\ to $\Omega_Xh^2$ as a function of $M_X/H_I$. The shaded
area is where thermalization {\em may} occur if the annihilation cross
section is its maximum value.  Also shown is the contribution assuming
that the \WIMPZILLA\ is present at the end of inflation with a
temperature $T=H_I/2\pi$.}  
\leavevmode\epsfxsize=300pt \epsfbox{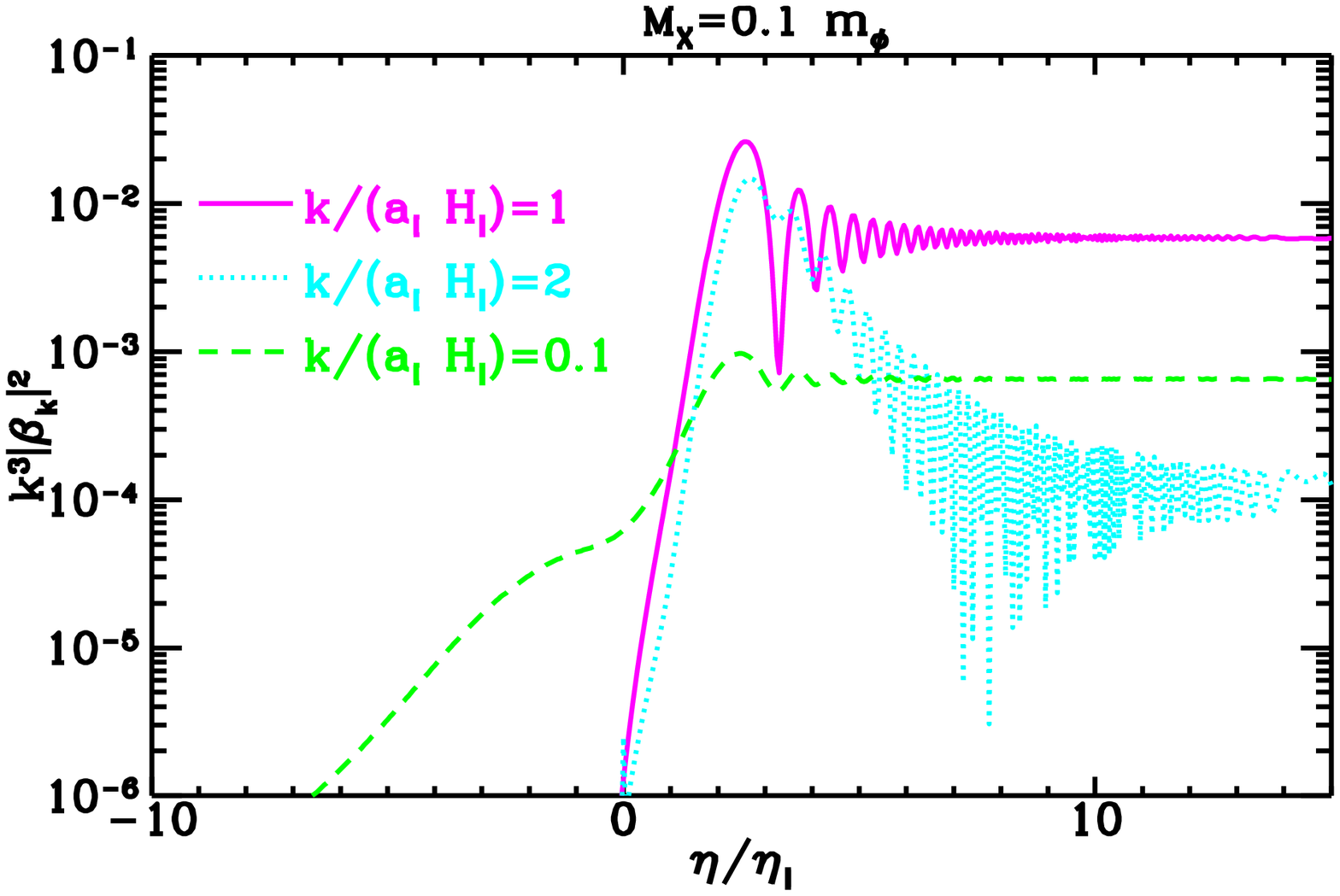}
\caption{\label{kofeta} The evolution of the  Bogoliubov coefficient with conformal 
time for several wavenumbers.  $\eta=\eta_I$ corresponds to the end of the
inflationary era.}
\end{figure}

Start with the canonical quantization of the $X$ field in an action of
the form (with metric $ds^2= dt^2- a^2(t) d{\bf x}^2 = a^2(\eta)
\left[d\eta^2 - d{\bf x}^2\right] $ where $\eta$ is conformal time)
\begin{equation}
S=\int dt \int d^3\!x\, \frac{a^3}{2}\left( \dot{X}^2 - \frac{(\nabla
X)^2}{a^2} - M_X^2 X^2 - \xi R X^2 \right)
\end{equation}
where $R$ is the Ricci scalar.  After transforming to conformal time
coordinate, use the mode expansion
\begin{equation}
X({\bf x})=\int \frac{d^3\!k}{(2 \pi)^{3/2} a(\eta)} \left[a_k h_k(\eta) e^{i
{\bf{k \cdot x}}} + a_k^\dagger h_k^*(\eta) e^{-i {\bf{k \cdot x}}}\right],
\end{equation}
where because the creation and annihilation operators obey the
commutator $[a_{k_1}, a_{k_2}^\dagger] = \delta^{(3)}({\bf k}_1 -{\bf
k}_2)$, the $h_k$s obey a normalization condition $h_k h_k^{'*} - h_k'
h_k^* = i$ to satisfy the canonical field commutators (henceforth, all
primes on functions of $\eta$ refer to derivatives with respect to
$\eta$).  The resulting mode equation is
\begin{equation}
h_k''(\eta) + w_k^2(\eta) h_k(\eta) = 0,
\label{eq:modeequation}
\end{equation}
where 
\begin{equation}
w_k^2= k^2 + M_X^2 a^2 + (6 \xi - 1) a''/a \ .
\label{eq:frequency}
\end{equation}
The parameter $\xi$ is 1/6 for conformal coupling and 0 for minimal
coupling.  From now on, $\xi=1/6$ for simplicity but without much loss
of generality.  By a change in variable $\eta \rightarrow k/a$, one
can rewrite the differential equation such that it depends only on
$H(\eta)$, $H'(\eta)/k$, $k/a(\eta)$, and $M_X$.  Hence, the
parameters $H_I$ and $a_I$ correspond to the Hubble parameter and the
scale factor evaluated at an arbitrary conformal time $\eta_I$, which
can be taken to be the approximate time at which $X$s are produced
({\it i.e.,} $\eta_I$ is the conformal time at the end of inflation).

One may then rewrite Eq.\ \ref{eq:modeequation} as
\begin{equation}
h_\ktilde''(\etatilde) + \left(\ktilde^2 + 
\frac{M_X^2}{H_I^2} \tilde{a}^2\right) h_\ktilde(\etatilde) 
=0 \ ,
\label{eq:scaledmodeequation}
\end{equation}
where $\etatilde=\eta a_I H_I$, $\tilde{a}=a/a_I$, and $\ktilde=
k/(a_I H_I)$.  For simplicity of notation,  drop all the
tildes.  This differential equation can be solved once the
boundary conditions are supplied.

The number density of the \WIMPZILLAS\ is found by a Bogoliubov
transformation from the vacuum mode solution with the boundary
condition at $\eta=\eta_0$ (the initial time at which the vacuum of
the universe is determined) into the one with the boundary condition
at $\eta= \eta_1$ (any later time at which the particles are no longer
being created).  $\eta_0$ will be taken to be $-\infty$ while $\eta_1$
will be taken to be at $+\infty$.  Defining the Bogoliubov
transformation as $ h_k^{\eta_1}(\eta)= \alpha_k h_k^{\eta_0}(\eta) +
\beta_k h_k^{* \eta_0}(\eta)$ (the superscripts denote where the
boundary condition is set), the energy density of produced particles
is
\begin{equation}
 \rho_X(\eta_1) = M_X n_X(\eta_1) = M_X
H_I^3\left (\frac{1}{\tilde{a}(\eta_1)}\right)^3 \int_0^{\infty}
\frac{d\tilde{k}}{2
\pi^2} \tilde{k}^2 |\beta_{\tilde{k}}|^2, 
\end{equation} 
where one should note that the number operator is defined at $\eta_1$
while the quantum state (approximated to be the vacuum state) defined
at $\eta_0$ does not change in time in the Heisenberg representation.

As one can see from Eq.\ \ref{eq:scaledmodeequation}, the input
parameter is $M_X/H_I$.  One must also specify the behavior of
$a(\eta)$ near the end of inflation.  In Fig.\ \ref{hawking} (from
\cite{grav}), I show the resulting values of $\Omega_Xh^2$ as a
function of $M_X/H_I$ assuming the evolution of the scale factor
smoothly interpolates between exponential expansion during inflation
and either a matter-dominated universe or radiation-dominated
universe.  The peak at $M_X/H_I \sim 1$ is similar to the case
presented in Ref.\ \cite{birrelldavies1980}.  As expected, for large
$M_X/H_I$, the number density falls off faster than any inverse power
of $M_X/H_I$.

Now most of the action occurs around the transition from inflation to
the matter-dominated or radiation-dominated universe.  This is shown
in Fig.\ \ref{kofeta}.  Also from Fig.\ \ref{kofeta} one can see that
most of the particles are created with wavenumber of order $H_I$.

To conclude, there is a significant mass range ($0.1 H_I$ to $H_I$,
where $H_I \sim 10^{13}$GeV) for which \WIMPZILLAS\ will have critical
density today regardless of the fine details of the transition out of
inflation.  Because this production mechanism is inherent in the
dynamics between the classical gravitational field and a quantum
field, it needs no fine tuning of field couplings or any coupling to
the inflaton field.  However, only if the particles are stable (or
sufficiently long lived) will these particles give contribution of the
order of critical density.

\subsection{Production during Reheating}

Another attractive origin for \WIMPZILLAS\ is during the defrosting
phase after inflation.  It is important to recall that it is not
necessary to convert a significant fraction of the available energy
into massive particles; in fact, it must be an infinitesimal amount.
I now will discuss how particles of mass much greater than $T_{RH}$
may be created in the correct amount after inflation in reheating
\cite{reheating}.

In one extreme is the assumption that the vacuum energy of inflation
is immediately converted to radiation resulting in a reheat
temperature $T_{RH}$.  In this case $\Omega_X $ can be calculated by
integrating the Boltzmann equation with initial condition $N_X=0$ at
$T=T_{RH}$.  One expects the $X$ density to be suppressed by
$\exp(-2M_X/T_{RH})$; indeed, one finds $\Omega_X \sim 1$ for
$M_X/T_{RH} \sim 25 + 0.5\ln(m_X^2\langle \sigma |v|\rangle)$, in
agreement with previous estimates \cite{vadimvaleri} that for
$T_{RH}\sim10^9$GeV, the \WIMPZILLA\ mass would be about
$2.5\times10^{10}$GeV.

A second (and more plausible) scenario is that reheating is not
instantaneous, but is the result of the slow decay of the inflaton
field.  The simplest way to envision this process is if the comoving
energy density in the zero mode of the inflaton decays into normal
particles, which then scatter and thermalize to form a thermal
background.  It is usually assumed that the decay width of this
process is the same as the decay width of a free inflaton field.

There are two reasons to suspect that the inflaton decay width might
be small.  The requisite flatness of the inflaton potential suggests a
weak coupling of the inflaton field to other fields since the
potential is renormalized by the inflaton coupling to other fields
\cite{review2}.  However, this restriction may be evaded in
supersymmetric theories where the nonrenormalization theorem ensures a
cancelation between fields and their superpartners.  A second reason
to suspect weak coupling is that in local supersymmetric theories
gravitinos are produced during reheating.  Unless reheating is
delayed, gravitinos will be overproduced, leading to a large undesired
entropy production when they decay after big-bang nucleosynthesis
\cite{ellis}.

It is simple to calculate the \WIMPZILLA\ abundance in the slow
reheating scenario.  It will be important to keep in mind that what is
commonly called the reheat temperature, $T_{RH}$, is not the maximum
temperature obtained after inflation.  The maximum temperature is, in
fact, much larger than $T_{RH}$.  The reheat temperature is best
regarded as the temperature below which the universe expands as a
radiation-dominated universe, with the scale factor decreasing as
$g_*^{-1/3}T^{-1}$.  In this regard it has a limited meaning
\cite{book,turner}.  One implication of this is that it is incorrect
to assume that the maximum abundance of a massive particle species
produced after inflation is suppressed by a factor of
$\exp(-M/T_{RH})$.

To estimate \WIMPZILLA\ production in reheating, consider a model
universe with three components: inflaton field energy, $\rho_\phi$,
radiation energy density, $\rho_R$, and \WIMPZILLA\ energy density,
$\rho_X$.  Assume that the decay rate of the inflaton field energy
density is $\Gamma_\phi$.  Also assume the \WIMPZILLA\ lifetime is
longer than any timescale in the problem (in fact it must be longer
than the present age of the universe).  Finally, assume that the light
degrees of freedom are in local thermodynamic equilibrium.

With the above assumptions, the Boltzmann equations describing the
redshift and interchange in the energy density among the different
components is
\begin{eqnarray}
\label{eq:BOLTZMANN}
& &\dot{\rho}_\phi + 3H\rho_\phi +\Gamma_\phi\rho_\phi = 0
	\nonumber \\
& & \dot{\rho}_R + 4H\rho_R - \Gamma_\phi\rho_\phi
   - 	\frac{\langle\sigma|v|\rangle}{m_X}
	\left[ \rho_X^2 - \left( \rho_X^{EQ} \right)^2 \right] =0
\nonumber \\
& & \dot{\rho}_X + 3H\rho_X 
    + \frac{\langle\sigma|v|\rangle}{m_X}
	\left[ \rho_X^2 - \left( \rho_X^{EQ} \right)^2 \right] = 0  \ ,
\end{eqnarray}
where dot denotes time derivative.  As already mentioned, $\langle
\sigma|v| \rangle$ is the thermal average of the $X$ annihilation
cross section times the M{\o}ller flux factor.  The equilibrium energy
density for the $X$ particles, $\rho_X^{EQ}$, is determined by the
radiation temperature, $T=(30\rho_R/\pi^2g_*)^{1/4}$.

It is useful to introduce two dimensionless constants, $\alpha_\phi$
and $\alpha_X$, defined in terms of $\Gamma_\phi$ and $\langle \sigma
|v| \rangle$ as
\begin{equation}
\label{alphagamma}
\Gamma_\phi = \alpha_\phi M_\phi \qquad
\langle \sigma |v| \rangle = \alpha_X M_X^{-2} \ .
\end{equation}
For a reheat temperature much smaller than $M_\phi$, $\Gamma_\phi$
must be small.  From Eq.\ (\ref{eq:TRH}), the reheat temperature in
terms of $\alpha_X$ and $M_X$ is $T_{RH}\simeq \alpha_\phi^{1/2}
\sqrt{M_\phi M_{Pl}}$.  For $M_\phi=10^{13}$GeV, $\alpha_\phi$ must be
smaller than of order $10^{-13}$.  On the other hand, $\alpha_X$ may
be as large as of order unity, or it may be small also.

It is also convenient to work with dimensionless quantities that can
absorb the effect of expansion of the universe.  This may be
accomplished with the definitions
\begin{equation}
\label{def}
\Phi \equiv \rho_\phi M_\phi^{-1} a^3 \ ; \quad
R    \equiv \rho_R a^4 \ ; \quad
X    \equiv \rho_X M_X^{-1} a^3 \ .
\end{equation}
It is also convenient to use the scale factor, rather than time, for
the independent variable, so one may define a variable $x = a M_\phi$.
With this choice the system of equations can be written as (prime
denotes $d/dx$)
\begin{eqnarray}
\label{eq:SYS}
\Phi' & = & - c_1 \ \frac{x}{\sqrt{\Phi x + R}}   \ \Phi \nonumber \\
R'    & = &   c_1 \ \frac{x^2}{\sqrt{\Phi x + R}} \ \Phi \
            + c_2 \ \frac{x^{-1}}{ \sqrt{\Phi x +R}} \
	           	         \left( X^2 - X_{EQ}^2 \right) \nonumber \\
X'    & = & - c_3 \ \frac{x^{-2}}{\sqrt{\Phi x +R}} \
		\left( X^2 - X_{EQ}^2 \right) \ .
\end{eqnarray}
The constants $c_1$, $c_2$, and $c_3$ are given by
\begin{equation}
c_1 = \sqrt{\frac{3}{8\pi}} \frac{M_{Pl}}{M_\phi}\alpha_\phi \ \qquad
c_2 = c_1\frac{M_\phi}{M_X}\frac{\alpha_X}{\alpha_\phi} \ \qquad
c_3 = c_2 \frac{M_\phi}{M_X} \ .
\end{equation}
$X_{EQ}$ is the equilibrium value of $X$, given in terms of the
temperature $T$ as (assuming a single degree of freedom for the $X$
species)
\begin{equation}
X_{EQ} = \frac{M_X^3}{M_\phi^3}\left( \frac{1}{2\pi} \right)^{3/2}
	x^3 \left(\frac{T}{M_X}\right)^{3/2}\exp(-M_X/T) \ .
\end{equation}
The temperature depends upon $R$ and $g_*$, the effective number of
degrees of freedom in the radiation:
\begin{equation}
\frac{T}{M_X} = \left( \frac{30}{g_*\pi^2}\right)^{1/4}
\frac{M_\phi}{M_X} \frac{R^{1/4}}{x} \ .
\end{equation}

It is straightforward to solve the system of equations in Eq.\
(\ref{eq:SYS}) with initial conditions at $x=x_I$ of $R(x_I)=X(x_I)=0$
and $\Phi(x_I)=\Phi_I$.  It is convenient to express
$\rho_\phi(x=x_I)$ in terms of the expansion rate at $x_I$, which
leads to
\begin{equation}
\Phi_I = \frac{3}{8\pi} \frac{M^2_{Pl}}{M_\phi^2}
		\frac{H_I^2}{M_\phi^2}\ x_I^3 \ .
\end{equation}
The numerical value of $x_I$ is irrelevant.

Before numerically solving the system of equations, it is useful to
consider the early-time solution for $R$.  Here, early times means $H
\gg \Gamma_\phi$, {\it i.e.,} before a significant fraction of the
comoving coherent energy density is converted to radiation.  At early
times $\Phi \simeq \Phi_I$, and $R\simeq X \simeq 0$, so the equation
for $R'$ becomes $R' = c_1 x^{3/2} \Phi_I^{1/2}$.  Thus, the early
time solution for $R$ is simple to obtain:
\begin{equation}
\label{eq:SMALLTIME}
R \simeq \frac{2}{5} c_1
     \left( x^{5/2} -  x_I^{5/2} \right) \Phi_I^{1/2}
			 \qquad (H \gg \Gamma_\phi) \ .
\end{equation}
Now express $T$ in terms of $R$ to yield the early-time
solution for $T$:
\begin{eqnarray}
\label{threeeights}
\frac{T}{M_\phi} & \simeq & \left(\frac{12}{\pi^2g_*}\right)^{1/4}
c_1^{1/4}\left(\frac{\Phi_I}{x_I^3}\right)^{1/8}  \nonumber \\
 & & \times	\left[ \left(\frac{x}{x_I}\right)^{-3/2} -
                \left(\frac{x}{x_I}\right)^{-4} \right]^{1/4}
		\qquad (H \gg \Gamma_\phi) \ .
\label{eq:approxtovmphi}
\end{eqnarray}
Thus, $T$ has a maximum value of
\begin{eqnarray}
\frac{T_{MAX}}{M_\phi}& = & 0.77
   \left(\frac{12}{\pi^2g_*}\right)^{1/4} c_1^{1/4}
   \left(\frac{\Phi_I}{x_I^3}\right)^{1/8} \nonumber \\ & = & 0.77
   \alpha_\phi^{1/4}\left(\frac{9}{2\pi^3g_*}\right)^{1/4} \left(
   \frac{M_{Pl}^2H_I}{M_\phi^3}\right)^{1/4} \ ,
\end{eqnarray}
which is obtained at $x/x_I = (8/3)^{2/5} = 1.48$.  It is also
possible to express $\alpha_\phi$ in terms of $T_{RH}$ and obtain
\begin{equation}
\label{max}
\frac{T_{MAX}}{T_{RH}} = 0.77 \left(\frac{9}{5\pi^3g_*}\right)^{1/8}
		\left(\frac{H_I M_{Pl}}{T_{RH}^2}\right)^{1/4} \ .
\end{equation}

For an illustration, in the simplest model of chaotic inflation $H_I
\sim M_\phi$ with $M_\phi \simeq 10^{13}$GeV, which leads to
$T_{MAX}/T_{RH} \sim 10^3 (200/g_*)^{1/8}$ for $T_{RH} =
10^9$GeV.

We can see from Eq.\ (\ref{eq:SMALLTIME}) that for $x/x_I>1$, in the
early-time regime $T$ scales as $a^{-3/8}$, which implies that entropy
is created in the early-time regime \cite{turner}.  So if one is
producing a massive particle during reheating it is necessary to take
into account the fact that the maximum temperature is greater than
$T_{RH}$, and that during the early-time evolution, $T\propto
a^{-3/8}$.

\begin{figure}[t]
\centering
\leavevmode\epsfxsize=300pt  \epsfbox{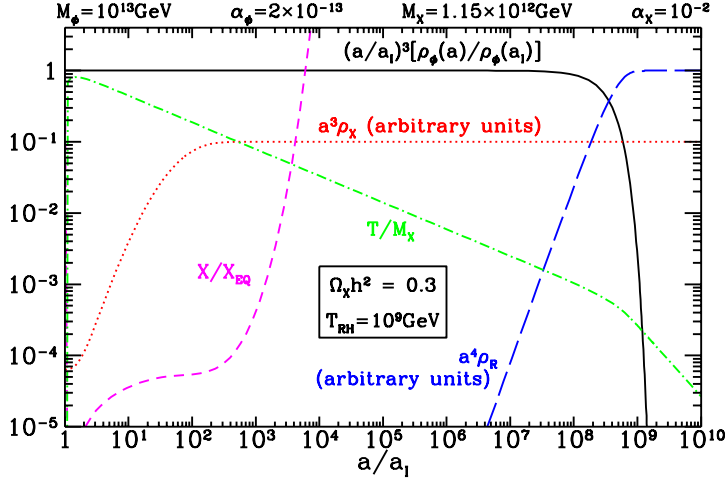}
\caption{\label{model1} The evolution of energy densities and $T/M_X$
as a function of the scale factor.  Also shown is $X/X_{EQ}$.}
\end{figure}

An example of a numerical evaluation of the complete system in Eq.\
(\ref{eq:SYS}) is shown in Fig.\ \ref{model1} (from \cite{reheating}).
The model parameters chosen were $M_\phi= 10^{13}$GeV, $\alpha_\phi
=2\times10^{-13} $, $M_X= 1.15\times10^{12}$GeV, $\alpha_X =10^{-2}$,
and $g_*=200$.  The expansion rate at the beginning of the coherent
oscillation period was chosen to be $H_I=M_\phi$.  These parameters
result in $T_{RH}=10^9$GeV and $\Omega_Xh^2=0.3$.

Figure \ref{model1} serves to illustrate several aspects of the
problem.  Just as expected, the comoving energy density of $\phi$
({\it i.e.}, $a^3\rho_\phi$) remains roughly constant until
$\Gamma_\phi\simeq H$, which for the chosen model parameters occurs
around $a/a_I\simeq 5\times10^8$.  But of course, that does not mean
that the temperature is zero.  Notice that the temperature peaks well
before ``reheating.''  The maximum temperature, $T_{MAX}= 10^{12}$GeV,
is reached at $a/a_I$ slightly larger than unity (in fact at
$a/a_I=1.48$ as expected), while the reheat temperature, $T_{RH}=
10^9$GeV, occurs much later, around $a/a_I\sim 10^8$.  Note that
$T_{MAX}\simeq 10^3 T_{RH}$ in agreement with Eq.\ (\ref{max}).

{}From the figure it is clear that $X \ll X_{EQ}$ at the epoch of
freeze out of the comoving $X$ number density, which occurs around
$a/a_I\simeq 10^2$.  The rapid rise of the ratio after freeze out is
simply a reflection of the fact that $X$ is constant while $X_{EQ}$
decreases exponentially.

A close examination of the behavior of $T$ shows that after the sharp
initial rise of the temperature, the temperature decreases as
$a^{-3/8}$ [as follows from Eq.\ (\ref{threeeights})] until $H\simeq
\Gamma_\phi$, and thereafter $T\propto a^{-1}$ as expected for the
radiation-dominated era.

For the choices of $M_\phi$, $\alpha_\phi$, $g_*$, and $\alpha_X$ used for
the model illustrated in Fig.\ \ref{model1}, $\Omega_Xh^2 = 0.3$ for  
$M_X=1.15\times10^{12}$GeV, in excellent agreement with the mass predicted 
by using an analytic estimate for the result \cite{reheating}
\begin{equation}
\label{om}
\Omega_X h^2 = M_X^2 \langle \sigma |v|\rangle \,
	\left(\frac{g_*}{200}\right)^{-3/2} \,
	\left (\frac{2000T_{RH}}{M_X}\right)^7 \ .
\end{equation}

Here again, the results have also important implications for the
conjecture that ultra-high cosmic rays, above the
Greisen-Zatsepin-Kuzmin cut-off of the cosmic ray spectrum, may be
produced in decays of superheavy long-living particles
\cite{dimitridark98,vadimvaleri,subircr,kr2}.  In order to produce
cosmic rays of energies larger than about $10^{13}$ GeV, the mass of
the $X$-particles must be very large, $M_X\simgt 10^{13}$ GeV and
their lifetime $\tau_X$ cannot be much smaller than the age of the
Universe, $\tau_X\simgt 10^{10}$ yr.  With the smallest value of the
lifetime, the observed flux of ultra-high energy cosmic rays will be
reproduced with a rather low density of $X$-particles, $\Omega_X\sim
10^{-12}$. It has been suggested that $X$-particles can be produced in
the right amount by usual collisions and decay processes taking place
during the reheating stage after inflation if the reheat temperature
never exceeded $M_X$ \cite{kr2}.  Again, assuming naively that that
the maximum number density of a massive particle species $X$ produced
after inflation is suppressed by a factor of $(M_X/T_{RH})^{3/2}
\exp(-M_X/T_{RH})$ with respect to the photon number density, one
concludes that the reheat temperature $T_{RH}$ should be in the range
$10^{11}$ to $10^{15}$GeV \cite{vadimvaleri}. This is a rather high
value and leads to the gravitino problem in generic supersymmetric
models.  This is one reason alternative production mechanisms of these
superheavy $X$-particles have been proposed
\cite{grav,kt,ckr2}. However, our analysis show that the situation is
much more promising. Making use of Eq.\ (\ref{om}), the right amount
of $X$-particles to explain the observed ultra-high energy cosmic rays
is produced for
\begin{equation}
\left(\frac{T_{RH}}{10^{10}\:{\rm GeV}}\right)\simeq
\left(\frac{g_*}{200}\right)^{3/14}\:\left(\frac{M_X}{10^{15}\:{\rm
GeV}}\right),
\end{equation}
where it has been assumed that $\langle \sigma |v|\rangle\sim
M_X^{-2}$. Therefore,  particles as massive as
$10^{15}$ GeV may be generated during the reheating stage in
abundances large enough to explain the ultra-high energy cosmic rays
even if the reheat temperature satisfies the gravitino bound.

\subsection{Production During Preheating}

Another way to produce \WIMPZILLAS\ after inflation is in a
preliminary stage of reheating called ``preheating''
\cite{preheating}, where nonlinear quantum effects may lead to an
extremely effective dissipational dynamics and explosive particle
production.

Particles can be created in a broad parametric resonance with a
fraction of the energy stored in the form of coherent inflaton
oscillations at the end of inflation released after only a dozen
oscillation periods.  A crucial observation for our discussion is that
particles with mass up to $10^{15}$ GeV may be created during
preheating \cite{kt,klr,gut}, and that their distribution is
nonthermal. If these particles are stable, they may be good candidates
for \WIMPZILLAS\ \cite{dan}.

The main ingredient of the preheating scenario introduced in the early
1990s is the nonperturbative resonant transfer of energy to particles
induced by the coherently oscillating inflaton fields.  It was
realized that this nonperturbative mechanism can be much more
efficient than the usual perturbative mechanism for certain parameter
ranges of the theory \cite{preheating}.  The basic picture can be seen
as follows.  Suppose there is a scalar field $X$ with a coupling $ g^2
\phi^2 X^2$ where $\phi$ is a homogeneous classical inflaton field.
The mode equation for $X$ field then can be written in terms of a
redefined variable $\chi_k \equiv X_k a^{3/2}$ as
\begin{equation} 
\ddot{\chi}_k (t) + [A + 2 q \cos(2t)] \chi_k(t)=0
\label{eq:mathieu}
\end{equation}
where $A$ depends on the energy of the particle and $q$ depends on the
inflaton field oscillation amplitude.  When $A$ and $q$ are constants,
this equation is usually referred to as the Mathieu equation which
exhibits resonant mode instability for certain values of $A$ and $q$.
In an expanding universe, $A$ and $q$ will vary in time, but if they
vary slowly compared to the frequency of oscillations, the effects of
resonance will remain.  If the mode occupation number for the $X$
particles is large, the number density per mode of the $X$ particles
will be proportional to $|\chi_k|^2$.  If $A$ and $q$ have the
appropriate values for resonance, $\chi_k$ will grow exponentially in
time, and hence the number density will attain an exponential
enhancement above the usual perturbative decay.  This period of
enhanced rate of energy transfer has been called preheating primarily
because the particles that are produced during this period have yet to
achieve thermal equilibrium.

This resonant amplification leads to an efficient transfer of energy
from the inflaton to other particles which may have stronger coupling to
other particles than the inflaton, thereby speeding up the reheating
process and leading to a higher reheating temperature than in the
usual scenario.  Another interesting feature is that particles of mass
larger than the inflaton mass can be produced through this coherent
resonant effect.  This has been exploited to construct a baryogenesis
scenario \cite{klr} in which the baryon number violating bosons with
masses larger than the inflaton mass are created through the resonance
mechanism.  A natural variation on this idea is to produce
\WIMPZILLAS\ by the same resonance mechanism.

Interestingly enough, what was found \cite{dan} is that in the context
of a slow-roll inflation with the potential $V(\phi)=m_\phi^2
\phi^2/2$ with the inflaton coupling of $g^2 \phi^2 X^2/2$, the
resonance phenomenon is mostly irrelevant to \WIMPZILLA\ production
because too many particles would be produced if the resonance is
effective.  For the tiny amount of energy conversion needed for
\WIMPZILLA\ production, the coupling $g^2$ must be small enough (for a
fixed $M_X$) such that the motion of the inflaton field at the
transition out of the inflationary phase generates just enough
nonadiabaticity in the mode frequency to produce \WIMPZILLAS\ .  The
rest of the oscillations, damped by the expansion of the universe,
will not contribute significantly to \WIMPZILLA\ production as in the
resonant case.  In other words, the quasi-periodicity necessary for a
true resonance phenomenon is not present in the case when only an
extremely tiny fraction of the energy density is converted into
\WIMPZILLAS.  Of course, if the energy scales are lowered such that a
fair fraction of the energy density can be converted to \WIMPZILLAS\
without overclosing the universe, this argument may not apply.

The main finding of a detailed treatment \cite{dan} is that
\WIMPZILLAS\ with a mass as large as $10^3 H_I$, where $H_I$ is the
value of the Hubble expansion rate at the end of inflation, can be
produced in sufficient abundance to be cosmologically significant
today.

If the \WIMPZILLA\ is coupled to the inflaton $\phi$ by a term
$g^2\phi^2X^2/2$, then the mode equation in Eq.\ \ref{eq:frequency} is
now changed to
\begin{equation}
\omega_k^2 + k^2 + \left(M_X^2 + g^2\phi^2\right)a^2 \ ,
\end{equation}
again taking $\xi = 1/6$.

The procedure to calculate the \WIMPZILLA\ density is the same
as in Section \ref{gravprod}.  Now, in addition to the parameter
$M_X/H_I$, there is another parameter $g M_{Pl}/H_I$.  Now
in large-field models $H_I\sim 10^{13}$GeV, so $M_{Pl}/H_I$
might be as large as $10^6$.  The choice of $g=10^{-3}$ would
yield $g M_{Pl}/H_I=10^3$.

\begin{figure}[p]
\centering
\leavevmode\epsfxsize=300pt  \epsfbox{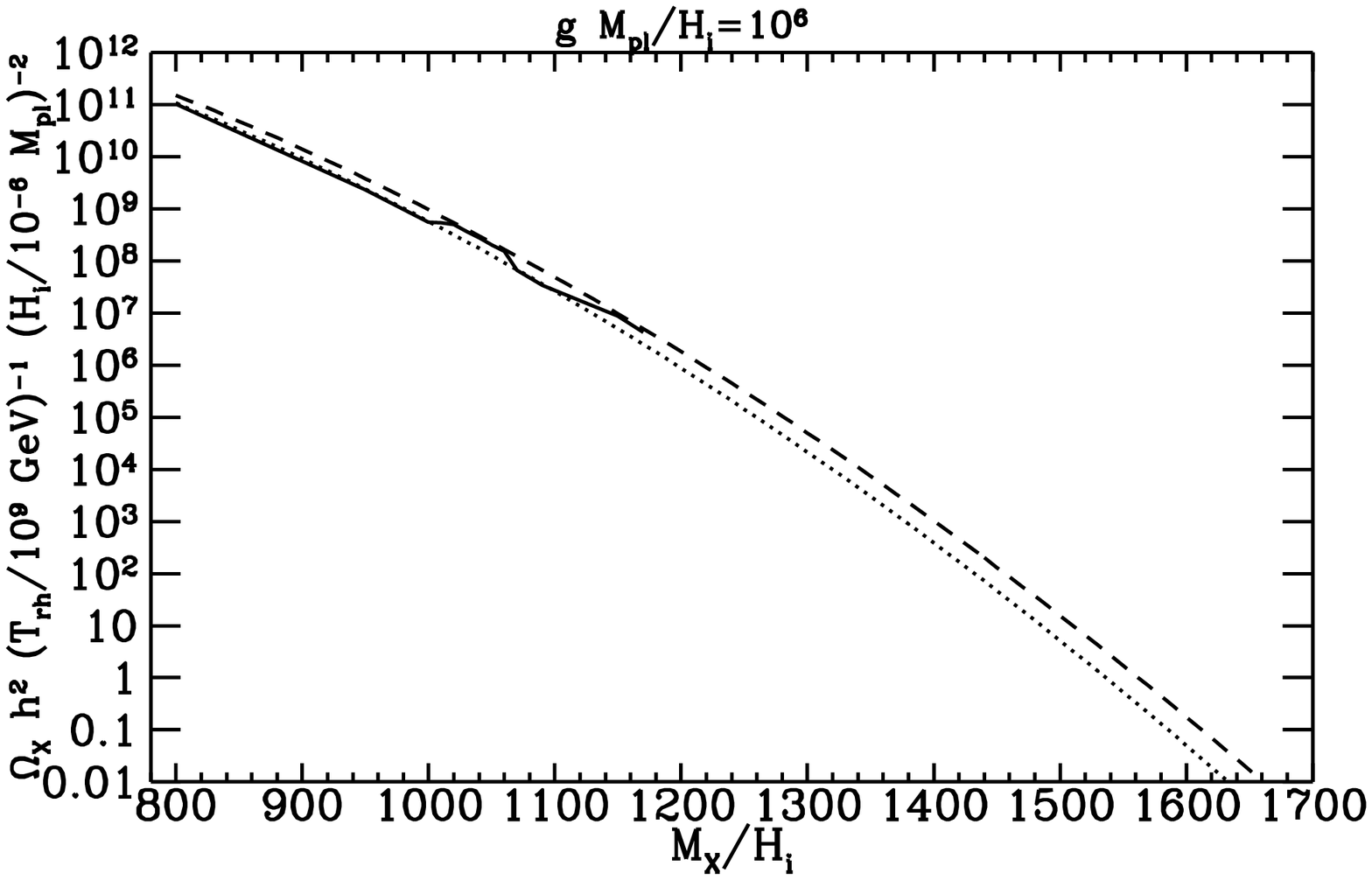} 
\caption{\label{dan1}A graph of $\Omega_Xh^2$ versus $M_X/H_I$ for
$gM_{Pl}/H_I= 10^6$.   The solid curve is a numerical result, while
the dashed and dotted curves are analytic approximations.}
\leavevmode\epsfxsize=300pt \epsfbox{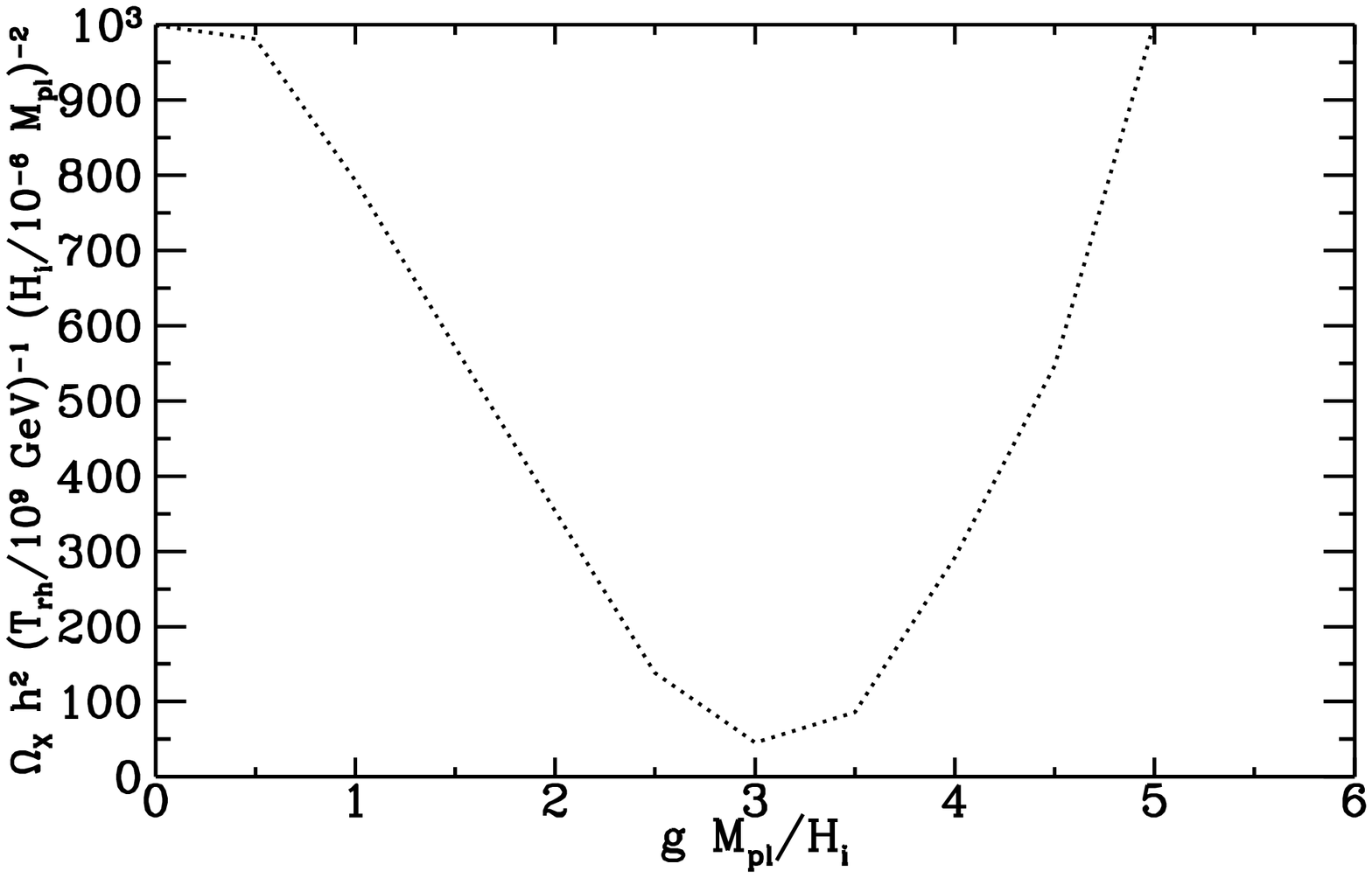}
\caption{\label{dan2} An illustration of the nonmonotonic behavior of
the particle density produced with the variation of the coupling
constant.  The value of $M_X/H_I$ is set to unity.}
\end{figure}

Fig.\ \ref{dan1} (from \cite{dan}) shows the dependence of the
\WIMPZILLA\ density upon $M_X/H_I$ for the particular choice $g
M_{Pl}/H_I=10^6$.  This would correspond to $g\sim 1$ in large-field
inflation models where $M_{Pl}/H_I=10^6$, about the largest possible
value.  Note that $\Omega_X\sim 1$ obtains for $
M_X/H_I\approx10^3$. The dashed and dotted curves are two analytic
approximations discussed in \cite{dan}, while the solid curve is the
numerical result.  The approximations are in very good agreement with
the numerical results.

Fig.\ \ref{dan2} (also from \cite{dan}) shows the dependence of the
\WIMPZILLA\ density upon $g M_{Pl}/H_I$.  For this graph $M_X/H_I$ was
chosen to be unity.  This figure illustrates the fact that the
dependence of $\Omega_Xh^2$ on $g M_{Pl}/H_I$ is not monotonic.  For
details, see \cite{dan}.

\subsection{Production in Bubble Collisions}

\WIMPZILLAS\ may also be produced \cite{ckr2} in theories where
inflation is completed by a first-order phase transition \cite{ls}, in
which the universe exits from a false-vacuum state by bubble
nucleation \cite{guth}.  When bubbles of true vacuum form, the energy
of the false vacuum is entirely transformed into potential energy in
the bubble walls.  As the bubbles expand, more and more of their
energy becomes kinetic as the walls become highly relativistic.

In bubble collisions the walls oscillate through each other
\cite{moss} and their kinetic energy is dispersed into low-energy
scalar waves \cite{moss,wat}.  We are interested in the potential
energy of the walls, $M_P = 4\pi\eta R^2$, where $\eta$ is the energy
per unit area of a bubble wall of radius $R$.  The bubble walls can be
visualized as a coherent state of inflaton particles, so the typical
energy $E$ of the products of their decays is simply the inverse
thickness of the wall, $E\sim \Delta^{-1}$. If the bubble walls are
highly relativistic when they collide, there is the possibility of
quantum production of nonthermal particles with mass well above the
mass of the inflaton field, up to energy $\Delta^{-1}=\gamma M_\phi$,
with $\gamma$ the relativistic Lorentz factor.

Suppose for illustration that the \WIMPZILLA\ is a fermion coupled to
the inflaton field by a Yukawa coupling $g \phi\overline{X}{X}$. One
can treat $\phi$ (the bubbles or walls) as a classical, external field
and the \WIMPZILLA\ as a quantum field in the presence of this source.
The number of \WIMPZILLAS\ created in the collisions from the wall
potential energy is $N_X\sim f_X M_P/M_X$, where $f_X$ parametrizes
the fraction of the primary decay products in \WIMPZILLAS.  The
fraction $f_X$ will depend in general on the masses and the couplings
of a particular theory in question.  For the Yukawa coupling $g$, it
is $ f_X \simeq g^2 {\rm ln}\left(\gamma M_\phi/2 M_{X}\right)$
\cite{wat,mas}.  \WIMPZILLAS\ may be produced in bubble collisions out
of equilibrium and never attain chemical equilibrium. Even with
$T_{RH}$ as low as 100 GeV, the present \WIMPZILLA\ abundance would be
$\Omega_{X}\sim 1$ if $g\sim 10^{-5}\alpha^{1/2}$.  Here
$\alpha^{-1}\ll 1$ is the fraction of the bubble energy at nucleation
in the form of potential energy at the time of collision.  This simple
analysis indicates that the correct magnitude for the abundance of
\WIMPZILLAS\ may be naturally obtained in the process of reheating in
theories where inflation is terminated by bubble nucleation.

\section{Conclusions}

In this talk I have pointed out several ways to generate nonthermal
dark matter.  All of the methods can result in dark matter much more
massive than the feeble little weak-scale mass thermal relics.  The
nonthermal dark matter may be as massive as the GUT scale, truly in
the \WIMPZILLA\ range.

The mass scale of the \WIMPZILLAS\ is determined by the mass scale of
inflation, more exactly, the expansion rate of the universe at the end
of inflation.  For large-field inflation models, that mass scale is of
order $10^{13}$GeV.  For small-field inflation models, it may be less,
perhaps much less.

The mass scale of inflation may one day be measured!  In addition to
scalar density perturbations, tensor perturbations are produced in
inflation.  The tensor perturbations are directly proportional to the
expansion rate during inflation, so determination of a tensor
contribution to cosmic background radiation temperature fluctuations
would give the value of the expansion rate of the universe during
inflation and set the scale for the mass of the \WIMPZILLA.

Undoubtedly, other methods for \WIMPZILLA\ production will be developed.
But perhaps even with the present scenarios one should start to
investigate methods for \WIMPZILLA\ detection.  While wimpy \WIMPS\ must
be color singlets and electrically neutral, \WIMPZILLAS\ may be endowed
with color and electric charge.  This should open new avenues for
detection and exclusion of \WIMPZILLAS.

\begin{figure}[t]
\centering
\leavevmode\epsfxsize=200pt  \epsfbox{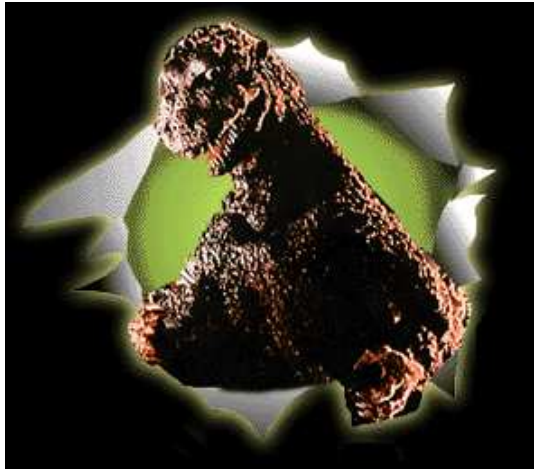}
\caption{\label{wimpzilla} Dark matter may be much more massive than
usually assumed, much more massive than wimpy \WIMPS, perhaps in the
\WIMPZILLA\ class.}
\end{figure}

The lesson of the talk is illustrated in Fig.\ \ref{wimpzilla}.
\WIMPZILLAS\ may surprise and be the dark matter, and we may learn
that size does matter!

\section*{Acknowledgements}
This work was supported by the Department of Energy and NASA (grant
number NAG5-7092).


\end{document}